\documentclass[11pt,a4paper]{article}
\pdfoutput=1

\usepackage{epsfig,latexsym,amsfonts,amsmath,amsthm,amssymb,amsbsy,multirow,slashed,color,mathrsfs,subfigure,wrapfig}
\usepackage[font={footnotesize,sl},bf]{caption}
\usepackage{jheppub}
\usepackage{mathtools}

\usepackage{etoolbox}
    \makeatletter
    \patchcmd{\maketitle}{\@fpheader}{}{}{}
    \makeatother

\usepackage{paralist}

\newcommand\be{\begin{equation}}
\newcommand\bea{\begin{eqnarray}}
\newcommand\ee{\end{equation}}
\newcommand\eea{\end{eqnarray}}

\newcommand{\bdm}{\begin{displaymath}}
\newcommand{\edm}{\end{displaymath}}

\newcommand{\f}[2]{\frac{#1}{#2}}
\newcommand{\bref}[1]{(\ref{#1})}
\newcommand\h{\frac{1}{2}}
\newcommand{\ket}[1]{|#1 \rangle}
\newcommand{\bra}[1]{\langle #1 |}
\newcommand{\defeq}{\vcentcolon=}

\title{Firewalls in AdS/CFT}
\author[1]{Steven G.\ Avery}
\author[2]{and Borun D.\ Chowdhury}
\affiliation[1]
{The Institute of Mathematical Sciences,\\
CIT Campus, Taramani, Chennai, India 600113}
\affiliation[2]
{Institute for Theoretical Physics, University of Amsterdam,\\
Science Park 904, Postbus 94485, 1090 GL Amsterdam, The Netherlands}

\abstract{Several recent papers argue against
  firewalls by relaxing the requirement for locality outside the stretched
  horizon. In the firewall argument, locality essentially serves the purpose of
  ensuring that the degrees of freedom required for infall are those in the
  proximity of the black hole and not the ones in the early radiation. We make
  the firewall argument sharper by utilizing the AdS/CFT framework and claim
  that the firewall argument essentially states that the dual to a thermal state
  in the CFT is a firewall.  }

\begin{document}
    \hspace{1in} 
\maketitle

\section{Introduction and summary}

The firewall phenomenon~\cite{Almheiri:2012rt} has reignited interest in the
information paradox.  Almheiri, Marolf, Polchinski and Sully (AMPS) have argued
that the postulates of black hole complementarity as stated in
\cite{Susskind:1993if,Susskind:1993mu} (including an implicit assumption of
smoothness of the horizon for an infalling observer) are mutually
inconsistent. Explicitly, unitarity of black hole evaporation and validity of
semi-classical physics outside the stretched horizon imply that observers
falling through an ``old" black hole horizon see intense radiation. This has
fueled a passionate debate, with most papers contesting the
\emph{firewall}~\cite{Nomura:2012sw,Mathur:2012jk,Banks:2012nn,Brustein:2012jn,Avery:2012tf,Larjo:2012jt,Rama:2012fm,Papadodimas:2012aq,Nomura:2012ex,Hsu:2013cw,Giddings:2012gc,Giddings:2013kcj,Jacobson:2012gh,Harlow:2013uq}. However,
some papers have also come out in
support~\cite{Bousso:2012as,Susskind:2012rm,Susskind:2012uw}, and some have been
of a clarifying and commenting
nature~\cite{Hossenfelder:2012mr,Chowdhury:2012vd,Bena:2012zi,Giveon:2012kp,Ori:2012jx,Hwang:2012nn,Page:2012zc,Saravani:2012is,Susskind:2013kx,Page:2013dx}.

A line of thought that challenges the firewall phenomenon gives up the validity
of local semi-classical physics outside the horizon to arbitrary distances. This
argument has come to be known as $\mathcal C=\mathcal A$.\footnote{Note that $\mathcal
  C=\mathcal A$ is not the only possible kind of non-locality one can
  consider in this context, cf.~\cite{Giddings:2012gc, Giddings:2013kcj}.} While
the notation will become clear in the bulk of the text, the argument essentially
says that the degrees of freedom inside the horizon are a subset of the degrees
of freedom of the early radiation far away from the black hole. Papers proposing
this view
include~\cite{Bousso:2012as,Papadodimas:2012aq,Jacobson:2012gh,Harlow:2013uq}. This
is manifestly non-local, but there are good reasons to expect non-locality in
quantum gravity; however, it would be interesting to quantify how much
non-locality is needed and see if it is reasonable.\footnote{For instance,
  reasonable might mean non-locality that can be attributed to effects within
  string theory.}  An attempt to put the $\mathcal C=\mathcal A$ idea on a
stronger footing is made in~\cite{Harlow:2013uq} by claiming that extracting the
degrees of freedom in the early hawking radiation responsible for free infall is
computationally impossible before the black hole evaporates.

Our agnosticism towards locality in quantum gravity, or rather an atheism towards it, makes the above resolution rather appealing; however, to check the reasonableness of
this idea, in this article we make the firewall argument more precise by
utilizing the AdS/CFT duality. We begin by looking at the evaporating
D1-D5-P black string. We repeat the firewall argument in this case, noting that after
the Page time the black string is highly entangled with the Hawking radiation
outside. The advantage of this system is that in certain limits this can be
viewed as excitations on the D1-D5 branes entangled with the radiation
outside. Since the D1-D5 system flows in the infrared to a CFT, in the
decoupling limit the near-horizon geometry can be viewed as a thermal state in
the CFT.

Taking the next logical step, we let an arbitrary system play the role of the
early radiation. In other words, we imagine coupling a source/sink to the CFT,
allowing them to equilibrate and thereby become entangled, and then decouple
them. Next, we couple a source to the CFT to create an infalling observer. The
$\mathcal C=\mathcal A$ argument in this case would mean that the degrees of
freedom of the source/sink that purifies the CFT are available to the infalling
observer, allowing her free infall. Since the systems are decoupled, this seems
to be a bizarre state of affairs given that we are talking about arbitrary
(decoupled) systems giving universal free infall. We discuss the implications of this and suggest that the dual
to a thermal state in the CFT is a firewall!

\section{The evaporating D1-D5 system and firewalls} \label{EvaporatingD1D5}

The full backreacted non-extremal D1-D5-P system in flat spacetime is a black string, whose
near-horizon geometry is the BTZ black hole~\cite{Banados:1992wn}. The full geometric solution may
be viewed as interpolating between BTZ and flat spacetime. The black string evaporates by Hawking
radiation and thus the recent blackhole--firewall--fuzzball debate can be embedded in this system.
The advantage being that the near-horizon region is dual to the D1-D5 CFT deformed by
irrelevant deformations that couple it to the flat space~\cite{David:1998ev,Avery:2009tu}. We can then
understand the implications of the firewall argument within AdS/CFT.

\begin{figure}[htbp]
\begin{center}
\subfigure[]{
\includegraphics[scale=.15]{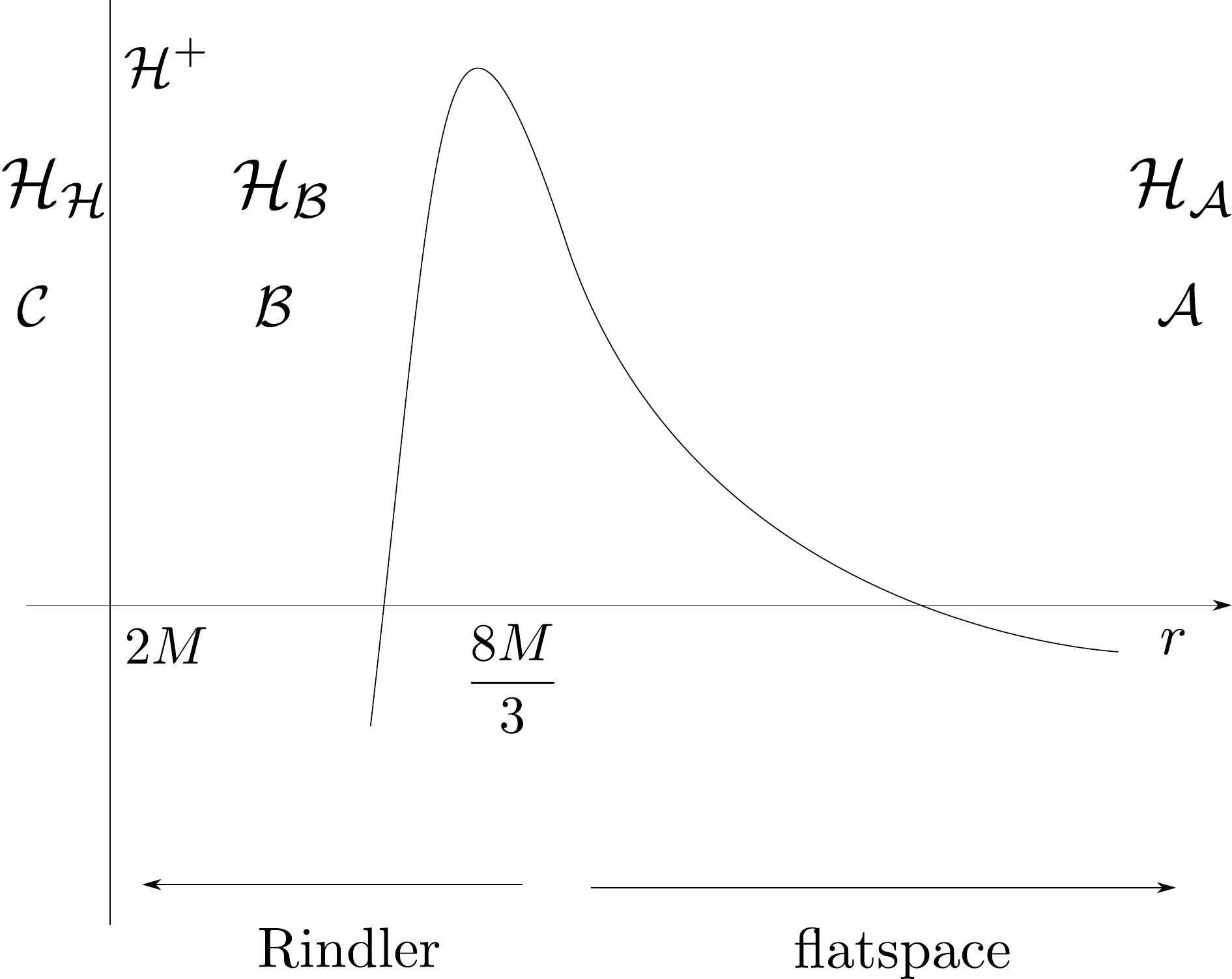}} \hspace{1in}
\subfigure[]{ 
\includegraphics[scale=.15]{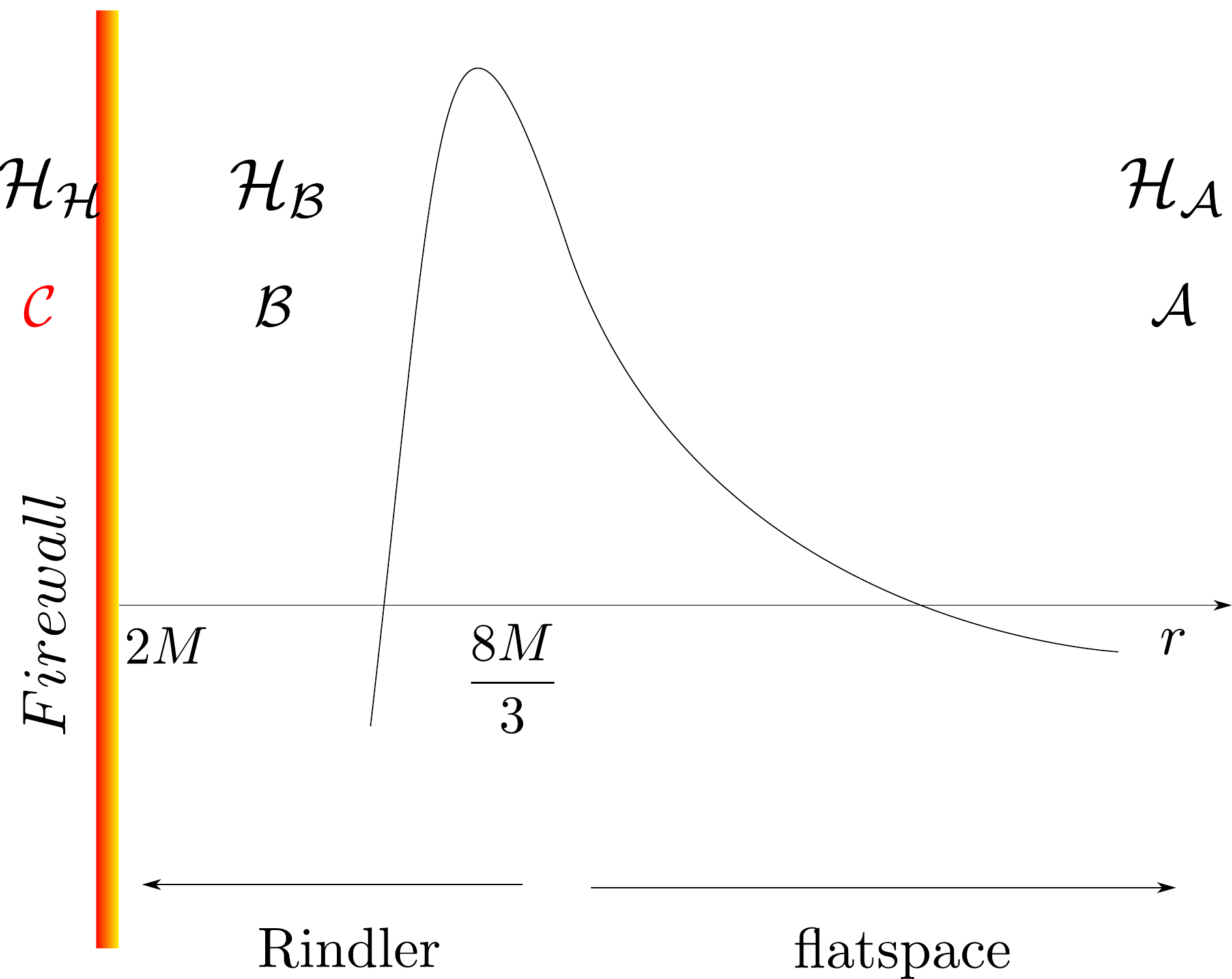} } 
\caption{In (a) a Schwarzschild black hole with the effective potential for a
  minimally coupled scalar is shown. The asymptotic flat space, its associated
  Hilbert space $\mathcal H_{\mathcal A}$ and modes living in it $\mathcal A$
  are on the outside of the potential barrier.  The near-horizon region, its
  associated Hilbert space $\mathcal H_{\mathcal B}$ and associated modes
  $\mathcal B$ are between the horizon and the barrier. Finally there is the
  region inside the horizon which is not accessible in Schwarzschild coordinates
  but is in Kruskal coordinates (for instance). Black hole complementarity
  posits that the experiences of an asymptotic and an infalling observer are
  complementary. The inside of the black hole is replaced by a stretched horizon
  for the asymptotic observer. Assuming the inside and stretched horizon Hilbert
  spaces to be isomorphic we denote it by $\mathcal H_{\mathcal H}$ and the
  associated degrees of freedom by $\mathcal C$. Free infall requires $\mathcal
  B$ and $\mathcal C$ to be maximally entangled.  In (b) the firewall is shown
  which is supposed arise for old black holes because the entanglement structure
  of $\mathcal A$ and $\mathcal B$ preclude maximal entanglement between
  $\mathcal B$ and $\mathcal C$.  }
\label{SchwarzschildGrayBody}
\end{center}
\end{figure}

To begin, let us review the essential features of the firewall argument. We start with a
Schwarzschild black hole formed by the collapse of matter in a pure state. The near-horizon region
of the Schwarzschild black hole is the Rindler geometry. This region is separated from the asymptotic
flat spacetime by a potential barrier whose exact details depend on the probe. According to black
hole complementarity~\cite{Susskind:1993if,Susskind:1993mu}, asymptotic observers describe the
black hole as a hot membrane unitarily evaporating. For such an observer, the Hilbert space naturally
factorizes at any time into subfactors as 
\be
\mathcal H = \mathcal H_{\mathcal H} \otimes \mathcal H_{\mathcal B} \otimes \mathcal H_{\mathcal A},
\ee 
where the modes populated by Hawking radiation that have escaped to flat space live in
$\mathcal{H}_{\mathcal A}$, modes inside the barrier but outside the horizon live in $\mathcal
H_{\mathcal B}$, and the Hilbert space associated with the stretched horizon degrees of freedom is
denoted by $\mathcal H_{\mathcal H}$. This is depicted in Figure~\ref{SchwarzschildGrayBody}a. For a
freely falling observer to pass through the horizon unscathed, one requires the state to be either the
Rindler or the Hartle--Hawking vacuum. Both of these have the modes across the horizon maximally
entangled with each other.\footnote{In fact, it is only the low-energy modes that are maximally
  entangled since there is a finite effective temperature. We abuse the term ``maximally
  entangled'' in this sense throughout our discussion. One can think of it as meaning
  maximally entangled \emph{given} conservation of energy.} Implicitly assuming that the inside of the
horizon is constructed from degrees of freedom in $\mathcal H_{\mathcal H}$~\cite{Susskind:2013kx},
AMPS conclude that free infall requires that the modes $\mathcal B \in \mathcal H_{\mathcal B}$ and
the modes $\mathcal C \in \mathcal H_{\mathcal H}$ be maximally entangled with each other. But after
the black hole has evaporated away half its entropy (i.e. after the Page
time~\cite{Page:1993df,Page:1993up,Sen:1996ph}), $\mathcal B$ has to be maximally entangled with the
early radiation $\mathcal A \in \mathcal H_{\mathcal A}$ in order to ensure unitarity. The monogamy
of entanglement then precludes $\mathcal B$ from being maximally entangled with $\mathcal C$. (This
is basically the contrapositive of Mathur's theorem against small corrections to the evaporation
process restoring unitarity~\cite{Mathur:2009hf,Avery:2011nb}.) AMPS emphasize that this means an infalling
observer cannot freely pass through the horizon after the Page time. Largely agreeing with AMPS,
Susskind argues for firewalls in a slightly different way in~\cite{Susskind:2012uw}. After the Page
time, the system inside the potential barrier is maximally entangled with the outside system and
thus the system inside the barrier cannot be split into two maximally entangled parts (viz.
$\mathcal B$ and $\mathcal C$); this implies that there is no space inside the horizon. There is
another argument advanced by AMPS: once more than half the entropy of the black hole
has been radiated away, the radiation is the bigger part of the full system, which is in a pure
state. The infalling observer can then perform a very non-local and fine-grained, but viable,
measurement on the early radiation to project the state at the horizon to a state that is not the
Unruh vacuum. Since the horizon is highly red-shifted, any state other than the Unruh vacuum or the
Hartle--Hawking vacuum will be very hot for the infalling observer and hence this conjectured
phenomenon has been dubbed a firewall.

We now turn to framing the AMPS argument in the D1-D5 system. Consider type IIB compactified on
$S^1 \times T^4$ with the volume of $S^1$ given by $2 \pi R$ and the volume of $T^4$ given by $(2
\pi)^4 V$. The torus is taken to be string size. We wrap $n_1$ D1 branes on $S^1$ and $n_5$ D5
branes on $S^1 \times T^4$. This system has a ground state degeneracy of $2 \sqrt{2} \pi \sqrt{n_1
  n_5}$~\cite{Balasubramanian:2005qu} which is accounted for by the Lunin--Mathur
geometries~\cite{Lunin:2002qf,Rychkov:2005ji}. We may think of starting with any one of these
states/geometries.

\begin{figure}[htbp]
\begin{center}
\includegraphics[scale=.25]{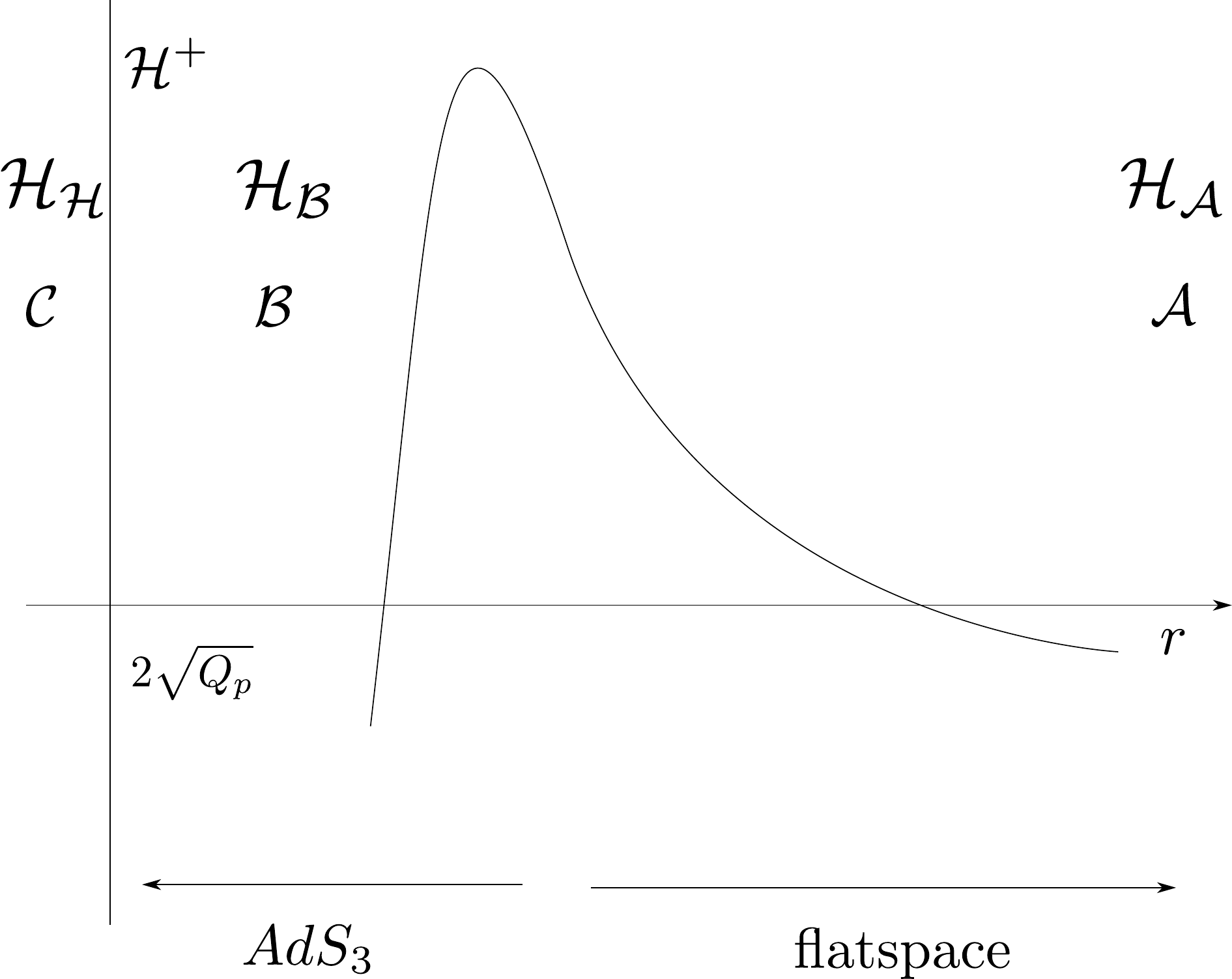}
\caption{The D1-D5-P black string with the effective potential barrier
  separating the flat space from the near horizon BTZ region. The Hilbert spaces
  and associated degrees of freedom have the same interpretation as in Figure
  \ref{SchwarzschildGrayBody}. For the traditional horizon with free
  infall, $\mathcal B$ and $\mathcal C$ have to be maximally entangled. }
\label{GrayBody}
\end{center}
\end{figure}

We can then make a black string by throwing in matter in the form of closed strings into these
geometries. We take the initial closed string state to be pure. In general any energy above
extremality excites all kinds of branes and anti-branes and momenta in all possible directions
consistent with equipartition of energy~\cite{Chowdhury:2006pk}. By taking the size of $S^1$ to be
much larger than $T^4$, however, the momentum along the $S^1$ becomes much lighter than any other
charges ensuring that it is preferentially excited. It is in this limit that the near-horizon region becomes 
$AdS_3$~\cite{Larsen:1999uk}. For simplicity, we consider the extra matter coming in with no net
momentum along the $S^1$, so that it then excites equal numbers of left and right movers. The metric for
this simplified system is\footnote{For the most general solution, see for
  example~\cite{Giusto:2004id,Cvetic:1996xz}.} 
\be 
ds^2 = \f{1}{\sqrt{H_1 H_5}} \left(-(1- \f{4Q_p}{r^2})dt^2 + dy^2 \right) 
         + \sqrt{H_1 H_5} \left(\f{dr^2}{1- \f{4 Q_p}{r^2}} + r^2 d\Omega_3^2 \right) 
         + \sqrt{\f{H_1}{H_5}} \sum_{i=1}^4 dx_i^2 
\ee 
where 
\be 
H_{1,5} = 1+ \f{Q_{1,5}}{r^2} 
\ee 
and 
\be
Q_1=n_1 g_s l_s^2, \qquad Q_5 = n_5 \f{g_s l_s^6}{V}, \qquad Q_P = n_p \f{g_s^2 l_s^8}{R^2 V};
\ee
$n_p$ is the left and right momentum measured in units of $R^{-1}$. The ADM mass of the black
string is 
\be 
M_{ADM} = \f{\pi}{4 G_5} \left[ Q_1 + Q_5 + 2 Q_p \right]. 
\ee 
where $16 \pi G_{10} = (2\pi)^7 g_s^2 l_s^8$ and $G_5 = \f{G_{10}}{(2 \pi)^5 RV}$. The core region of this geometry is $BTZ \times S^3 \times T^4$ with $R_{AdS_3} = R_{S^3} =
\sqrt{Q_1 Q_5}$, which is obtained by zooming into the region $r^2 \ll Q_1,Q_5$. This core region is
separated from the asymptotically flat region by probe-dependent potential barriers as shown in
Figure~\ref{GrayBody}. The horizon, near-horizon, and asymptotically flat Hilbert spaces are again
represented by $\mathcal H_{\mathcal H},\mathcal H_{\mathcal B}$ and $\mathcal H_{\mathcal A}$,
respectively.  The core region is dual to a $1+1$-dimensional $\mathcal N=(4,4)$ CFT, whose Hilbert
space, according to the AdS/CFT
duality~\cite{Aharony:1999ti}, is dual to $\mathcal
H_{\mathcal H} \otimes \mathcal H_{\mathcal B}$. Note the similarity of
Figures~\ref{SchwarzschildGrayBody}a and~\ref{GrayBody} but also note that in the Schwarzschild case
there is no decoupling limit, and no obvious gauge--gravity duality.

The left and right sectors of the D1-D5 CFT are on equal footing for our simplified system, with left and right moving momenta, entropy, and temperature:
\begin{gather}
P_{L,R} = \f{n_p}{R} \label{Momentum}, \\
S_{L,R} = 2 \pi \sqrt{n_1 n_5 n_p} \label{Entropy}, \\
T_{L,R} = \f{1}{R} \sqrt{\f{n_p}{n_1 n_5}}. \label{Temperature} 
\end{gather}

We are working in the dilute gas limit, $Q_p \ll Q_1,Q_5$, when the evaporation rate obtained from the bulk and the D1-D5 field theory match~\cite{Callan:1996dv,Dhar:1996vu,Das:1996wn,Das:1996jy,Maldacena:1996ix},
\be
\Gamma = 2 \pi^2 Q_1 Q_5 \f{\pi \omega}{2} \f{1}{e^{\omega/2 T_L}-1} \f{1}{e^{\omega/2 T_R}-1} \f{d^4k}{(2\pi)^4}.
\ee
For our system, this relation gives
\bea
\f{dn_p}{dt} \propto- g_s^2 l_s^4 \f{1}{R^5} \f{n_p^3}{(n_1n_5)^2}.
\eea
After time 
\be
t_\text{Page} \propto \f{R^5 (n_1 n_5)^2}{g_s^2 l_s^4 n_p^2}, \label{PageTime}
\ee
the system evaporates away half its entropy.\footnote{By taking $n_p \gg n_1 n_5$ we can ignore the entropy coming from ground state degeneracy of the D1-D5 system.}

\begin{figure}[htbp]
\begin{center}
\subfigure[]{
\includegraphics[scale=.3]{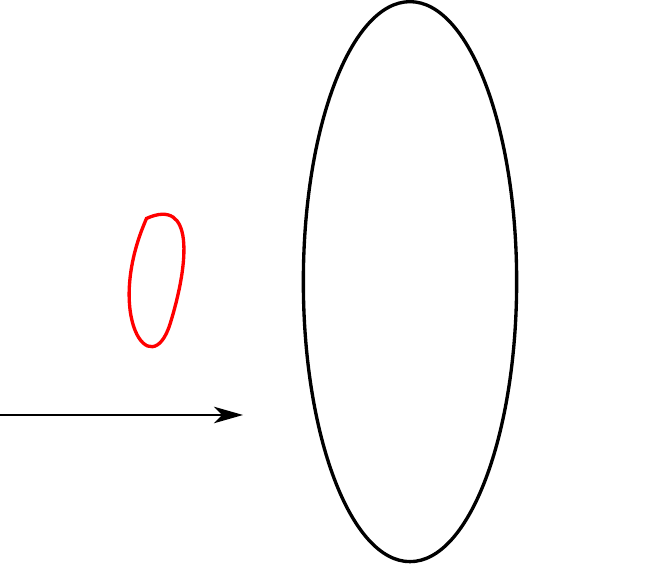}} \hspace{1in}
\subfigure[]{ 
\includegraphics[scale=.3]{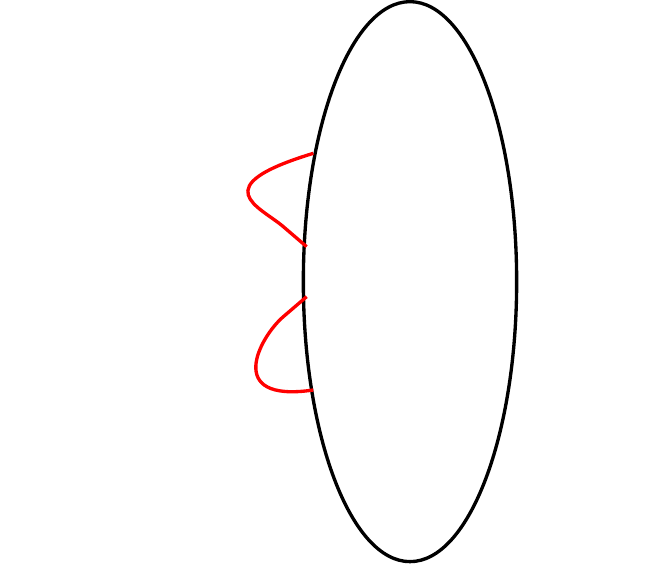} }  \\
\subfigure[]{
\includegraphics[scale=.3]{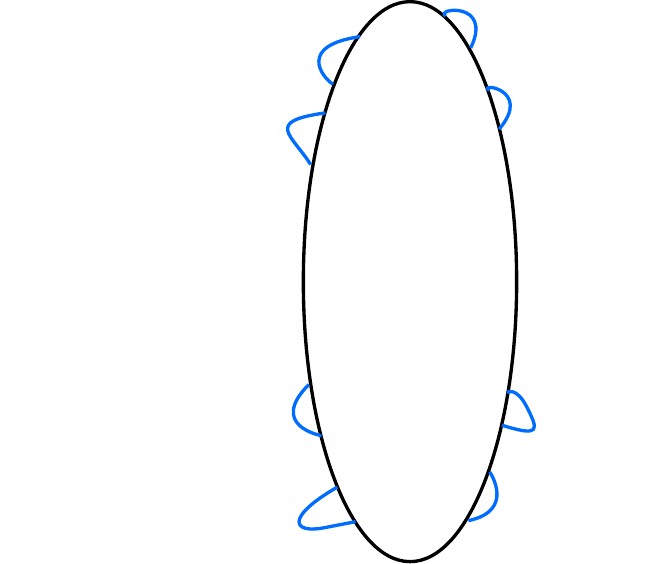}} \hspace{1in}
\subfigure[]{ 
\includegraphics[scale=.3]{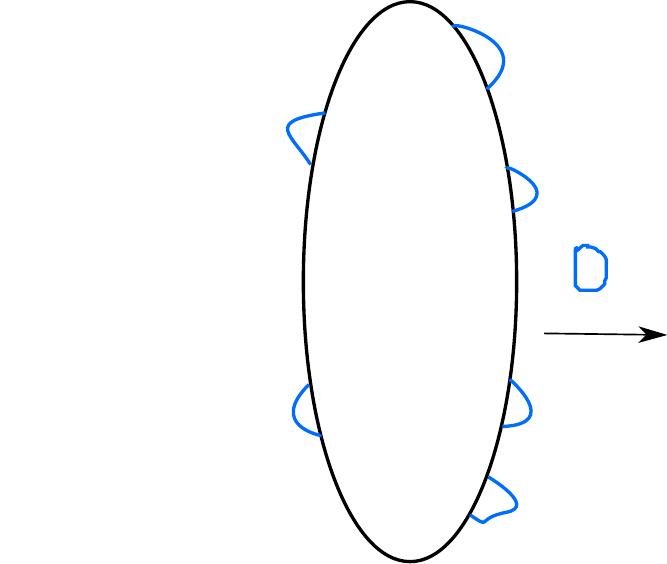} }  \\

\caption{Closed strings in a pure state hitting a stack of D-branes in (a)
  become open strings on the D-branes in (b). These open strings break into many
  lower energy open strings due to interactions on the D-branes in (c). These
  lower energy open strings then collide with each other and are emitted as
  closed strings because of the time reversal of the process (a) in (d). Since
  it is entropically unfavorable for all low-energy open strings to find one
  another at the same time so the same closed string as in (a) is not
  generically emitted. An effective arrow of time thus emerges.  }
\label{D-braneAdS}
\end{center}
\end{figure}

This may be interpreted as (a strong coupling version of) the  process shown in Figure
\ref{D-braneAdS}. Closed strings hit a stack of D1-D5 branes and become open strings  on them.
Fractionation of the branes~\cite{Dijkgraaf:1996xw,Mathur:2005ai} and the world-volume interactions
cause the open strings to break up into many lower energy open strings~\cite{Avery:2010vk,
  Avery:2010er, Avery:2010hs, Asplund:2011cq}. It is the
coarse-grained entropy of these excitations which account for the entropy of the D1-D5 CFT
\bref{Entropy}. With time these open strings collide and leave the D-branes as closed strings and
this process is interpreted as the dual of Hawking radiation. 
\begin{figure}[htbp]
\begin{center}
\includegraphics[scale=.25]{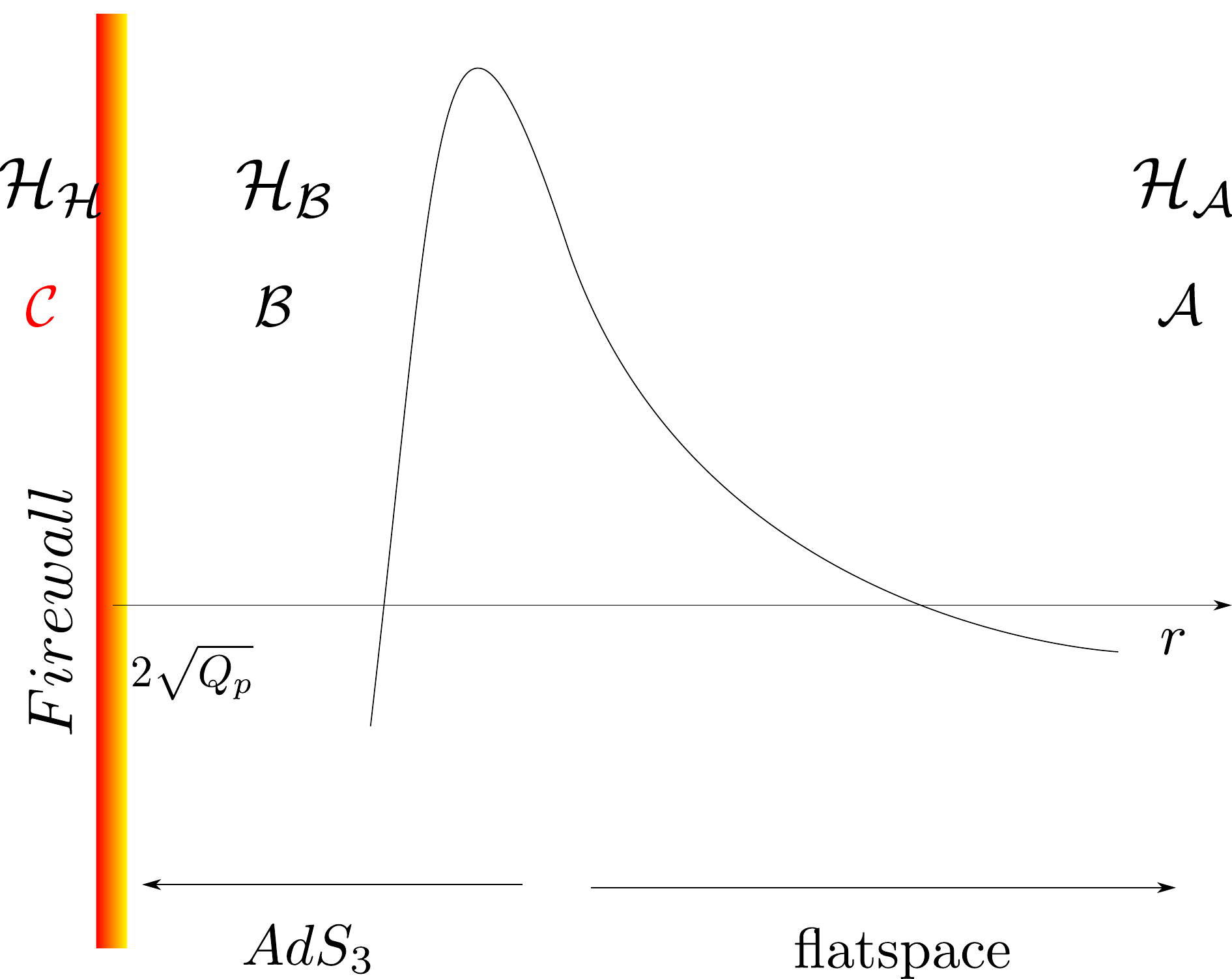}
\caption{The firewall for the D1-D5-P system is shown. The argument works just
  like that for the Schwarzschild case. The entanglement structure between
  $\mathcal A$ and $\mathcal B$ required by unitarity at late times precludes
  the entanglement structure between $\mathcal B$ and $\mathcal C$
  required for free infall. The added advantage in looking at the D1-D5-P system
  is that the near-horizon region is supposed to be dual to a CFT. We see that
  after the Page time the CFT is maximally entangled with early radiation
  $\mathcal A$.  }
\label{GrayBodyFirewall}
\end{center}
\end{figure}

The firewall argument can be made for the
D1-D5 system as follows. After the Page time~\bref{PageTime}, the brane system is maximally
entangled with the radiation outside. Following the line of reasoning in~\cite{Susskind:2012uw}, the core region dual to the brane
system now has a firewall instead of a harmless horizon. The picture before the Page
time is shown in Figure~\ref{GrayBody}, and the picture after the Page time according to the firewall
argument is shown in Figure~\ref{GrayBodyFirewall}.

The argument for the D1-D5-P black string runs just like the one for Schwarzschild black hole. What
do we gain by casting the firewall argument this way? The new feature here is that after Page time the
near-horizon region of this system is dual to the D1-D5 CFT in a thermal state. Thus, essentially
the firewall argument says that the dual to a thermal state\footnote{Due to leakage over the
  barrier the CFT has a physical cutoff; it is the state in the cutoff theory that is thermal. We
  believe this does not change the basic argument.}
is the firewall!

At this juncture, it is worth emphasizing a few points. First, it is obvious that this same
argument applies to other incarnations of AdS/CFT duality with an explicit brane construction.
Furthermore, let us note that there are several closely related physical scenarios to keep in mind: 
\begin{compactitem}
\item the near-horizon region of a very young black string, 
\item the near-horizon region of the post-Page time black string, 
\item the dual CFT at finite temperature, 
\item the CFT maximally entangled with a second CFT, 
\item and the CFT maximally entangled with some arbitrary heat bath. 
\end{compactitem}
All except for the first case are described by a thermal state in the CFT. The
first case is described by a (pure) typical state, which for many purposes can
be approximated by a thermal state.  Thus, according to the standard AdS/CFT
dictionary, all except possibly the first example, are expected to describe the
same asymptotically AdS geometry, which we just argued has a firewall. To escape
this conclusion, one must conjecture a generalization of superselection sectors
proposed in~\cite{Marolf:2012xe} as we discuss in the next
section. Alternatively, if one finds an evasion to the firewall argument, it may
be that none have a firewall; however, we want to emphasize that the
evasion better work for all of the above cases. Let us point out that if
firewalls form at the Page time and one does not postulate superselection
sectors, then observers may freely fall through the horizon only in the first
case, which is dual to a pure state in the CFT. It is amusing to note that this
is the opposite of the conclusion one might draw
from~\cite{VanRaamsdonk:2010pw,Czech:2012be,Czech:2012bh}.

The main point of this article is to make the preceding more precise and to address some of
the arguments against firewalls in light of this idea. We return to this after reviewing
these arguments.

\subsection{Two selected arguments against firewalls} \label{ArgumentsAgainstFirewalls}

While there have been many arguments against firewalls, as noted in the
Introduction, we review two that are especially relevant to this article. 

\subsubsection*{Papadodimas--Raju conjecture} 

In \cite{Papadodimas:2012aq}, Papadodimas and Raju argue that in the context of
AdS/CFT, infall is captured in the semi-classical limit by $n$-point functions
with $n \ll N$, the latter being the central charge of the CFT. It is claimed
the Hilbert space factorizes into a coarse-grained and fine-grained part
\be
\mathcal H = \mathcal H_\text{coarse} \otimes \mathcal H_\text{fine}. \label{HilbertSpaceFactor}
\ee
Since Papadodimas and Raju consider a large black hole in AdS that does not
evaporate, there is only what we refer to as $\mathcal B$ and $\mathcal C$
above, satisfying
 \be
\mathcal B \in \mathcal H_\text{coarse}, \qquad \mathcal C \in \mathcal H_\text{fine}. \label{PapadodimasRaju}
 \ee
In fact since there is no evaporation, there is no radiation, and
 \be
 \mathcal H_\text{coarse}=\mathcal H_{\mathcal B} , \qquad  \mathcal H_\text{fine}=\mathcal H_{\mathcal H};
 \ee
the fine-grained degrees of freedom account for the horizon degrees of freedom.
 
Even though the large black hole does not evaporate Papadodimas and Raju
conjecture that even when part of the fine-grained space has evaporated away
from the horizon, the degrees of freedom responsible for free fall inside the
horizon are the same as those in the radiation outside the horizon. In other
words, they claim that for an evaporating black hole
\be
 \mathcal H_\text{fine} = \mathcal H_{\mathcal H} \otimes \mathcal H_{\mathcal A}
\ee
and
\be
\mathcal C \in \mathcal H_{\mathcal H} \otimes \mathcal H_{\mathcal A},
\ee
so $\mathcal C$ can be found in either the fine-grained degrees of freedom
localized at the horizon or in the radiation spread over a typical distance
$t_\text{Page}$ away from it. As the black hole evaporates, it is increasingly
found in the radiation.  This opinion has also been suggested by
others~\cite{Bousso:2012as,Jacobson:2012gh,Harlow:2013uq} and is probably shared
by many others in unpublished form. This idea has come to be known as $\mathcal
C= \mathcal A$. This bypasses the firewall argument, which uses strong
subadditivity, by claiming that $\mathcal A$, $\mathcal B$ and $\mathcal C$ are
not independent systems. This still leaves the possibility that the infalling
observer (or someone else) may perform detailed experiments on $\mathcal A$ to
spoil infall and indeed such a possibility is acknowledged by Papadodimas and
Raju, but they claim that such an experiment is very hard to perform so
generically there will be free infall.

\subsubsection*{Harlow--Hayden conjecture}

In \cite{Harlow:2013uq}, Harlow and Hayden argue that the measurement on
$\mathcal A$ that AMPS showed would render the horizon a firewall, takes a
time $t \sim e^{M^2}$ while the time for the black hole to evaporate completely is
$t \sim M^3$. They conclude that it is not possible for an infalling observer
to perform the AMPS measurement before falling into the black
hole. Acknowledging that the argument using strong subadditivity does not
actually require the infalling observer to perform the measurement on the early
radiation, they propose the following criteria for breakdown of effective field
theory:
\begin{quote}
  Two spacelike-separated low-energy observables which are not both
  computationally accessible to some single observer do not need to be realized
  even approximately as distinct and commuting operators on the same Hilbert
  space.
\end{quote}
This computational complementarity proposal implies that the computational
complexity of measuring $\mathcal A$ in the way AMPS propose means that it is
possible that $\mathcal C$ has support on $\mathcal H_{\mathcal A}$, as was
suggested in \cite{Bousso:2012as,Papadodimas:2012aq,Jacobson:2012gh}, even
though they are spacelike-separated.\footnote{Harlow and Hayden go on to claim
  that a stronger form of complementarity is also consistent with their
  conjecture: the infalling observer's quantum mechanics may not
  be embeddable in that of the outside observer's. Namely, $\mathcal H_{\mathcal
    A}$ may not be part of the infalling observer's Hilbert space and then
  $\mathcal C$ may be maximally entangled with $\mathcal B$ for the infalling
  observer despite $\mathcal B$ being maximally entangled with $\mathcal A$ for
  the asymptotic observer. While we do not directly address the stronger form of
  complementarity in this paper, we would like to point out that it does not seem
  likely that $\mathcal H_{\mathcal A}$ may be missing from the infalling
  observer's quantum mechanics completely since the early radiation would
  backreact and influence the geodesic of the infalling observer.}

\section{Firewalls as duals to thermal CFT} \label{ThermalCFT}

In the previous section, we explain how an evaporating brane system becomes maximally
entangled with the radiation outside. The firewall argument can be
made for this system, and it implies that in the core region of the geometry,
the part that is dual to the low-energy limit of the branes, there is a
firewall.

However, we also review some rebuttals to the firewall argument, which state
that the degrees of freedom required for free infall, $\mathcal C$, do not only
come from the horizon degrees of freedom contained in $\mathcal H_{\mathcal H}$
but may have support in the radiation degrees of freedom in $\mathcal
H_{\mathcal A}$. This involves a certain degree of non-locality. The required
non-locality has been acknowledged in \cite{Papadodimas:2012aq,Harlow:2013uq},
but it is tolerated by saying we do not know enough about quantum gravity to
rule it out. Harlow and Hayden support their claim by noting that computational
complexity suggests that verification of the non-locality is not possible.

Let us understand what role locality plays in the firewall
argument. Locality mandates that the degrees of freedom required for free infall,
$\mathcal B$ and $\mathcal C$, are both present in the vicinity of the horizon
since that is where infall is taking place. While one's lack of understanding of
quantum gravity allows one to postulate that $\mathcal C$ may be present in
$\mathcal A$, we can make the puzzle sharper.

As we discuss above, the near-horizon region for the D1-D5 system is BTZ so we
can imagine decoupling the near-horizon region by making $S^1$ much larger than
all scales in the problem after the Page time. Alternatively, rather than using
the Hawking radiation to thermalize the branes, we can directly start with a CFT
in a thermal state. We can imagine that it was thermalized by coupling it to a
large source/sink with a Hilbert space $\mathcal H_{\mathcal S}$ that acted as a
heat bath. After equilibrium is attained, we then decouple $\mathcal{H}_{\mathcal
  S}$. Thus the state of the full system is 
\be 
\ket{\Psi} = \f{1}{\sqrt{Z}}
\sum_E ~ e^{-\beta E/2} ~ \ket{E}_{\mathcal H_{CFT}} \otimes \ket{E}_{\mathcal H_{\mathcal S}} \label{entangledState} 
\ee 
where $Z=\sum_E e^{-\beta E}$ and $\ket{E}_{\mathcal H_{CFT}}$ are states of the CFT and $\ket{E}_{\mathcal
  H_{\mathcal S}}$ are states of the heat bath. What is the dual to this thermal
state?

\begin{figure}[htbp]
\begin{center}
\subfigure[]{
\includegraphics[scale=.30]{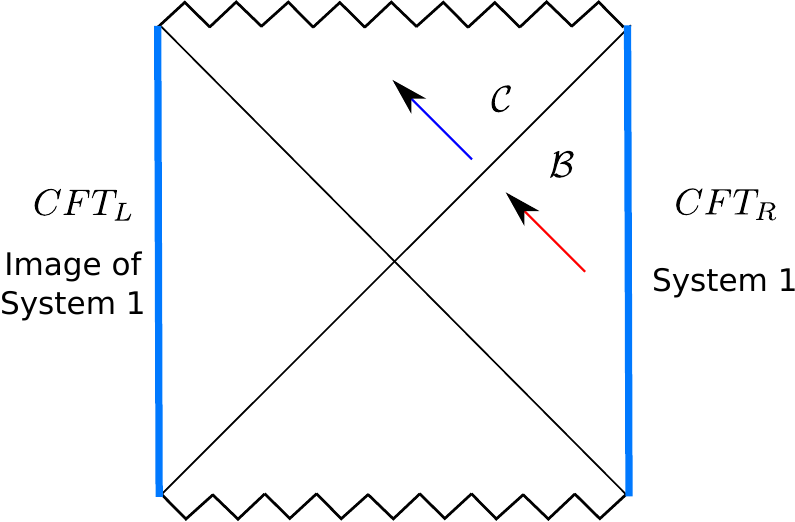}} \hspace{.5in}
\subfigure[]{
\includegraphics[scale=.12]{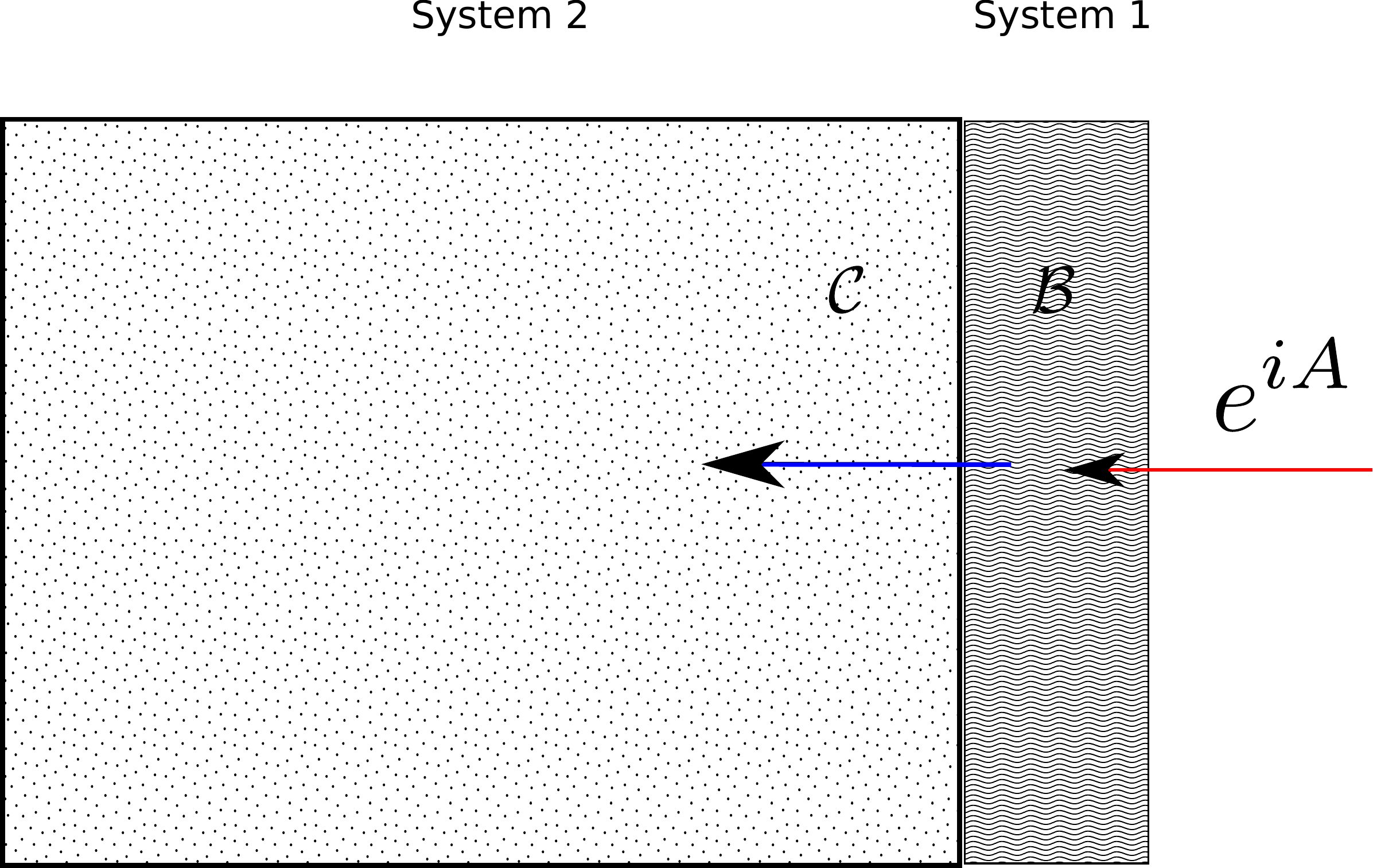}} 
\caption{(a) The eternal AdS black hole is dual to two CFTs entangled in a certain way. An excitation on the right side representing an infalling observer requires degrees of freedom associated with the left CFT to move past the horizon. In (b) this system is realized as System 1 being in contact and thermal equilibrium with a bigger System 2. 
}
\label{PureCFTInfall}
\end{center}
\end{figure}

In discussions, we have found many people claim that the dual of a thermal state
in the CFT is the eternal AdS black hole based on arguments
in~\cite{Maldacena:2001kr}.\footnote{Ref.~\cite{Maldacena:2001kr} proposes that the Lorentzian eternal $AdS_{d+1}$ black hole is dual to two decouples CFTs on $S^{d-1} \times R$ in the highly entangled thermofield double state~\bref{entangledState}. In this paper we assume this is true. For arguments refuting this proposal see~\cite{Avery:2013bea}.}  This is not correct.  In~\cite{Maldacena:2001kr}, it
is proposed that the dual description of maximally extended eternal AdS black
holes with two boundaries involves two CFTs living on the boundaries in an
entangled state resembling~\bref{entangledState} (with the Hilbert spaces being
those of the two boundary CFTs). As discussed in~\cite{Marolf:2012xe}, one can
create an infalling observer close to the right boundary of the geometry by
acting on $\text{CFT}_R$ with a unitary operator $e^{iA}$. Describing the
evolution of the observer past the horizon requires the degrees of freedom of
$\text{CFT}_L$. This is shown in Figure \ref{PureCFTInfall}. One realizes the
setup in the following way. The $\text{CFT}_R$ may describe some System 1 that
is in thermal equilibrium and in contact with a bigger System 2. Assuming it is
described by a conformal theory, the part of the bigger system that purifies the
smaller system may play the role of $\text{CFT}_L$. Note that an excitation created in
System 1 will eventually leak into System 2.

\begin{figure}[htbp]
\begin{center}
\subfigure[]{
\includegraphics[scale=.12]{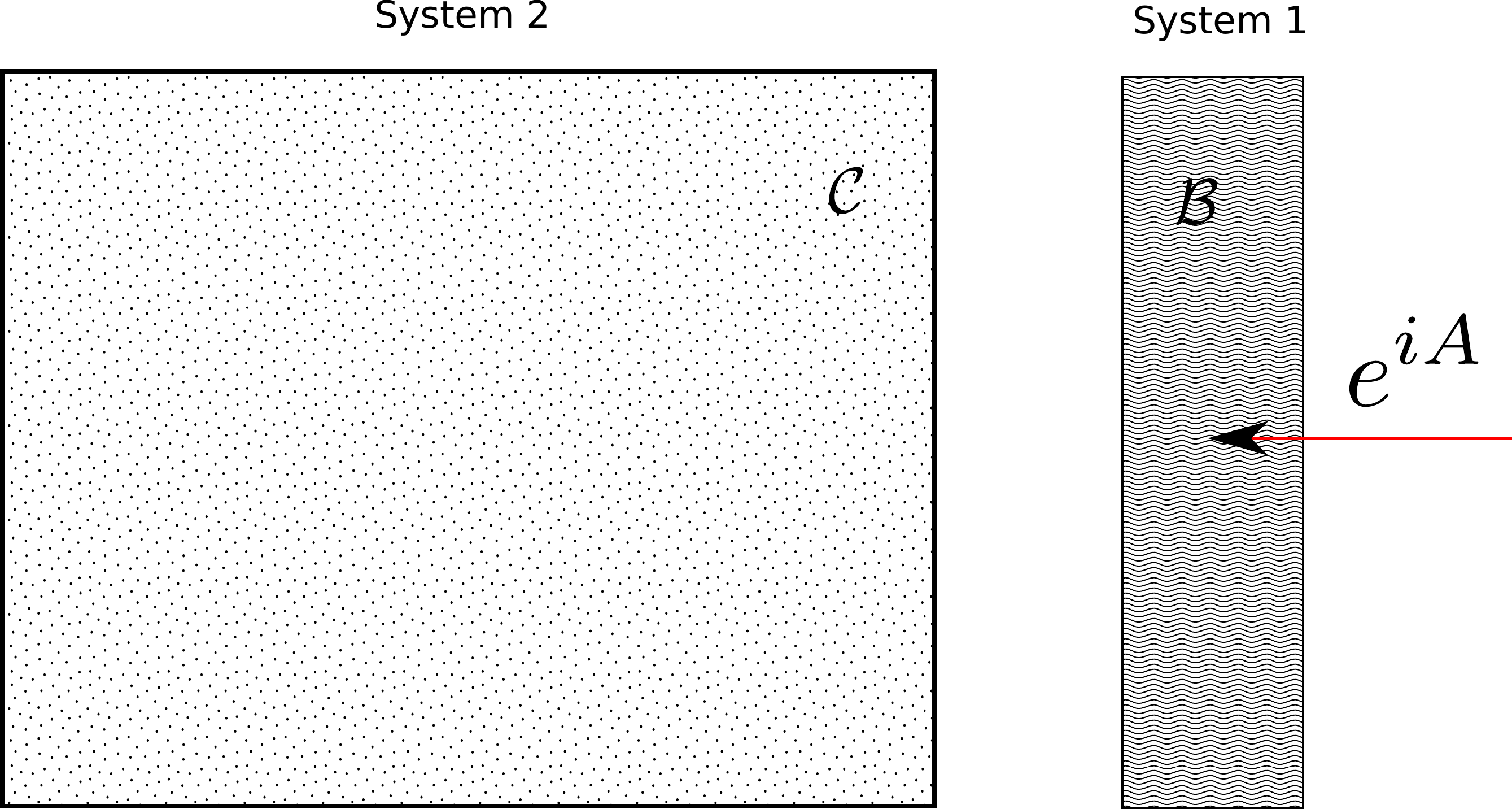}} \hspace{.5in}
\subfigure[]{
\includegraphics[scale=.30]{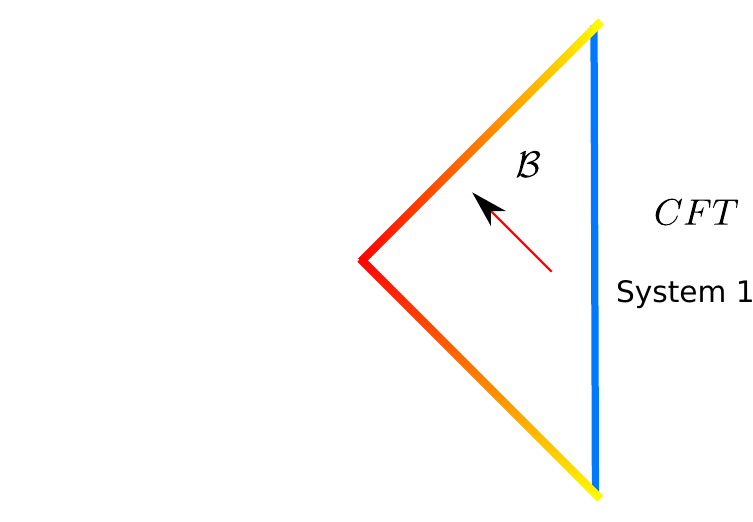}} 
\caption{We consider a CFT in a thermal state living on Hilbert space $\mathcal H_{CFT}$. We can imagine it was thermalized by a sink/source with degrees of freedom living on $\mathcal H_{\mathcal S}$. We take the two systems to be decoupled. Like the situation shown in Figure \ref{PureCFTInfall} we can consider infall of an observer coming from the right. However, unlike the situation in Figure \ref{PureCFTInfall}, the degrees of freedom of $\mathcal H_{\mathcal S}$ are not available and there is a firewall at the horizon.}
\label{ThermalCFTInfall}
\end{center}
\end{figure}

The situation we are interested in is subtly different. We are asking what is
the bulk dual of \emph{one} copy of the CFT, which is in a thermal state. We
simply do not have the other CFT's degrees of freedom that are necessary for
free infall. The equivalent of Papadodimas--Raju and Harlow--Hayden argument
would be that the degrees of freedom of $\mathcal H_{\mathcal S}$ (which is the
equivalent of $\mathcal H_{\mathcal A}$ for the evaporating branes) nevertheless
come into play. However, given that the we have decoupled the source/sink from the CFT and that the source/sink may not be a CFT or have a viable holographic description this seems rather implausible. The reason we say its implausible is that the crossing of a horizon involves a Bell measurement (a joint measurement) on the modes on either side. This means the observer crossing the horizon interacts with both set of modes. See Appendix~\ref{sec:bell} for more details. 
 It seems
rather implausible that an observer living on the CFT system would still be able
to access the degrees of freedom of $\mathcal H_{\mathcal S}$ in order to do a joint measurement,
irrespective of the properties of the latter system and independent of the coupling between the two systems. Said differently the evolution of a perturbation created with support on the CFT beyond the horizon depends not only on how the CFT is entangled with some other system but also on the Hamiltonian of the combined system.  We thus suggest that for
generic System 2 the infalling observer hits a firewall. This is shown in
Figure~\ref{ThermalCFTInfall}.\footnote{We should point out a possible loophole in the above reasoning. Joint measurement assumes the apparatus is coupled to both the systems during the decoherence process. Since the apparatus in this context lives on the CFT system we are inclined to say a joint measurement is not possible in the absence of a coupling between the two systems. However, the theory of decoherence is not very well understood particularly in the context of AdS/CFT. Without the same it is hard to completely rule out a reconciliation of the bulk and boundary measurement processes.}

The two cases---free infall as illustrated in Figure~\ref{PureCFTInfall} and
hitting a firewall as illustrated in Figure~\ref{ThermalCFTInfall}---are two
extremes. The answer to what is the dual to a thermal CFT seems beyond the
information in the one CFT. This can be seen as a generalization of
superselection sectors discussed in \cite{Marolf:2012xe}. How much of the
early radiation is available to act as the other copy of CFT for the stack of
evaporating D-branes is a question that can be rephrased as which superselection
sector quantum gravity is in. A question that seems can only be answered by
knowing the full theory of quantum gravity or by jumping into a black hole.

\begin{figure}[htbp]
\begin{center}
\includegraphics[scale=.12]{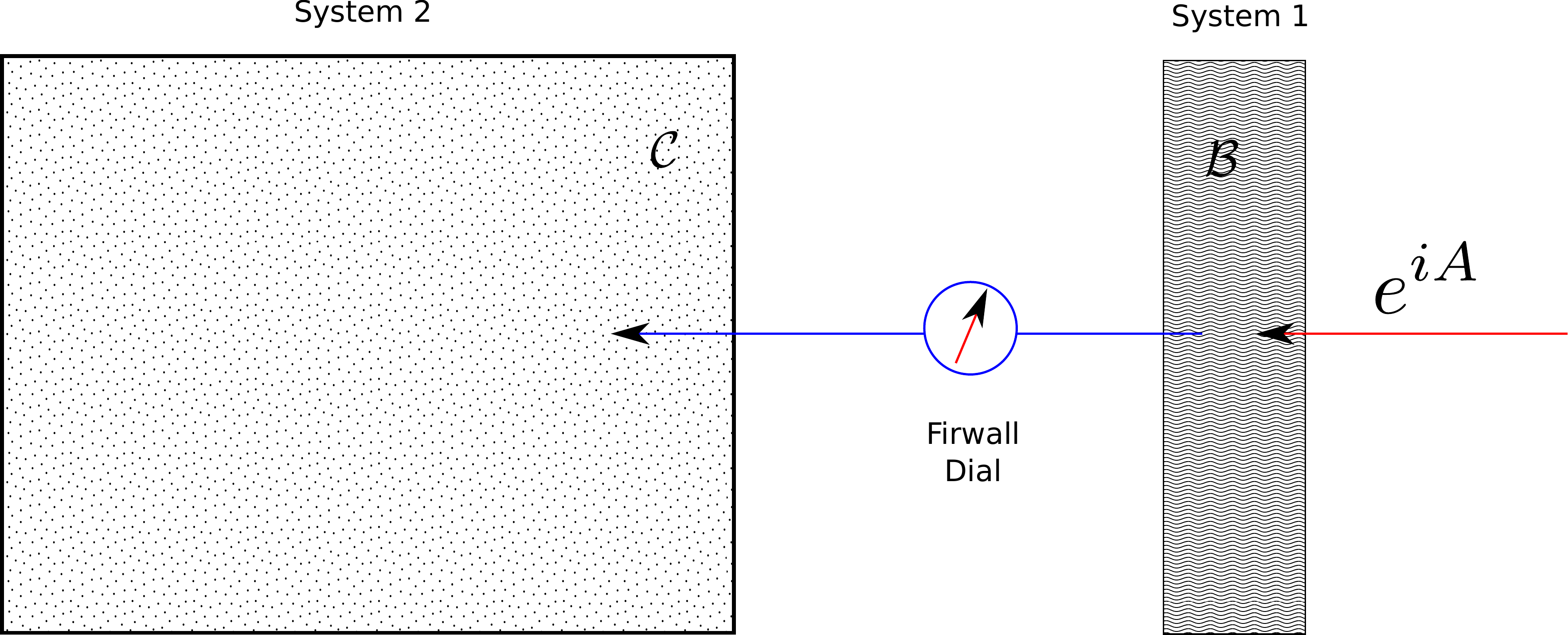}
\caption{How much and in what way the degrees of freedom purifying the thermal
  CFT are available for free infall is dependent on the full theory. We can also
  phrase this as free infall depends on which superselection sector quantum gravity lives in.}
\label{FirewallDial}
\end{center}
\end{figure}

\section*{Acknowledgements}

We would like to thank Jan de Boer, Bartek Czech, Daniel Harlow, Samir Mathur, Shiraz Minwalla,   Kyriakos Papadodimas, Ashoke Sen and Erik Verlinde   for helpful discussions. The work of BDC was supported by  the ERC Advanced Grant 268088-EMERGRAV.

\appendix

\section{Infall as Bell measurement}\label{sec:bell}

Immediately after AMPS's firewall paper~\cite{Almheiri:2012rt}, one of us proposed that the correct
way to analyse the situation would be think of the observer as part of the complete system and
measurements as coming from decoherence between the observer (or her apparatus) and the rest of the
system~\cite{Chowdhury:2012vd}. This appendix is based on the same theme and on the talk~\cite{michigan-talk}. 


Consider massless fields in $1+1$ dimensions. The equations of motion split the fields into left and
right movers. We consider only the left movers and the right movers behave the same. It can be shown
(see~\cite{Parentani:1993yz} for example) that the Minkowski vacuum can be expressed in terms of
Rindler modes as
\be
\ket{0_M} = \f{1}{\sqrt{\prod Z_\lambda}} \prod_\lambda e^{\tanh \theta_\lambda b_{\lambda,R}^\dagger b_{\lambda,L}^\dagger}\ket{0_R}\ket{0_L}
\ee
where $Z_\lambda = Tr[e^{-2 \pi \lambda/a}]$,  $\tanh \theta_\lambda =e^{-\pi \lambda/a}$ where $a$ is the acceleration of the Rindler observer. Different modes given by different $\lambda$ decouple and we can focus on the vacuum for a particular $\lambda$ 
\be
\ket{0_{M,\lambda}} = \f{1}{\sqrt{Z_\lambda}} \sum \tanh^n \theta_\lambda~ \ket{n_{\lambda,R}} \ket{n_{\lambda,L}}.
\ee
Note that if we consider the high temperature limit and restrict to fermionic modes then the above truncates to
\be
\ket{0_{M,\lambda}}  = \f{1}{\sqrt{2}} ( \ket{0_{\lambda,R}} \ket{0_{\lambda,L}} +  \ket{1_{\lambda,R}} \ket{1_{\lambda,L}} ) \label{MinkowskiMode}
\ee
and we can simplify our analysis by just considering qubits. The right moving observer will encounter left moving modes localised inside and outside the horizon and will find the state as the vacuum only if together they are in the state \bref{MinkowskiMode}. 

There is a simple generalization of the Minkwoski vacuum state~\bref{MinkowskiMode} which is a maximally entangled state between the two subsystems. One can write down four such orthogonal states
\begin{equation}\begin{aligned}
\ket{\varphi_1} &\defeq \frac{1}{\sqrt{2}}\big(\ket{\hat{0} }\ket{0 } 
                           + \ket{\hat{1} }\ket{1 }\big)\,,\\
\ket{\varphi_2} &\defeq \frac{1}{\sqrt{2}}\big(\ket{\hat{0} }\ket{0 } 
                           - \ket{\hat{1} }\ket{1 }\big)\,,\\
\ket{\varphi_3} &\defeq \frac{1}{\sqrt{2}}\big(\ket{\hat{0} }\ket{1 } 
                           + \ket{\hat{1} }\ket{0 }\big)\,,\\
\ket{\varphi_4} &\defeq \frac{1}{\sqrt{2}}\big(\ket{\hat{0} }\ket{1 } 
                           - \ket{\hat{1} }\ket{0 }\big)\,, \\ \label{BellStates}
\end{aligned}\end{equation}
and these are referred to as Bell states. The $\ket{\hat 0}$ and $\ket{\hat 1}$ are eigenstates of $\hat \sigma_z$ and similarly $\ket{0}$ and $\ket{1}$ are eigenstates of $\sigma_z$. Observe that in a simplified qubit model the Minkowski state corresponds to the first Bell state. 

The reduced density matrix of the hatted and the  unhatted systems for all four states are
\be
\hat \rho  = \h ( \ket{\hat 0} \bra{\hat 0} +  \ket{\hat 1} \bra{\hat 1} ), \qquad \rho  = \h ( \ket{0} \bra{0} +  \ket{1} \bra{1} )
\ee
which means that Charlie with access to only one of the systems (i.e. with access to operators $\hat I \otimes \sigma_x,~\hat I \otimes \sigma_y$ and $\hat I \otimes \sigma_z$) will get identical response from all four states and will be unable to distinguish between them. This does not, however, mean that the four states are indistinguishable. These states are eigenstates of the operators $\hat \sigma_x \otimes \sigma_x,~\hat \sigma_y \otimes \sigma_y$ and $\hat \sigma_z \otimes \sigma_z$. The eignevalues are shown in the table below.
\begin{center}
\begin{tabular}{|c|c|c|c|c|}
\hline
\text{state} & $\hat \sigma_x \otimes \sigma_x$ & $\hat \sigma_y \otimes \sigma_y$ & $\hat \sigma_z \otimes \sigma_z$ \\
\hline
\hline
$\ket{\varphi_1}$ & +1 & -1 & +1 \\
$\ket{\varphi_2}$ & -1 & +1 & +1 \\
$\ket{\varphi_3}$ & +1 & +1 & -1 \\
$\ket{\varphi_4}$ & -1 & -1 & -1 \\
\hline
\end{tabular} 
\end{center}
Thus, an observer, Alice, can distinguish between the four states by measuring the expectation value of any of the two operators, say $\hat \sigma_x \otimes \sigma_x$ and $\hat \sigma_z \otimes \sigma_z$. This is called a Bell measurement. 

\begin{figure}[htbp]
\begin{center}
\subfigure[Non-Bell Measurement]{
\includegraphics[scale=.35]{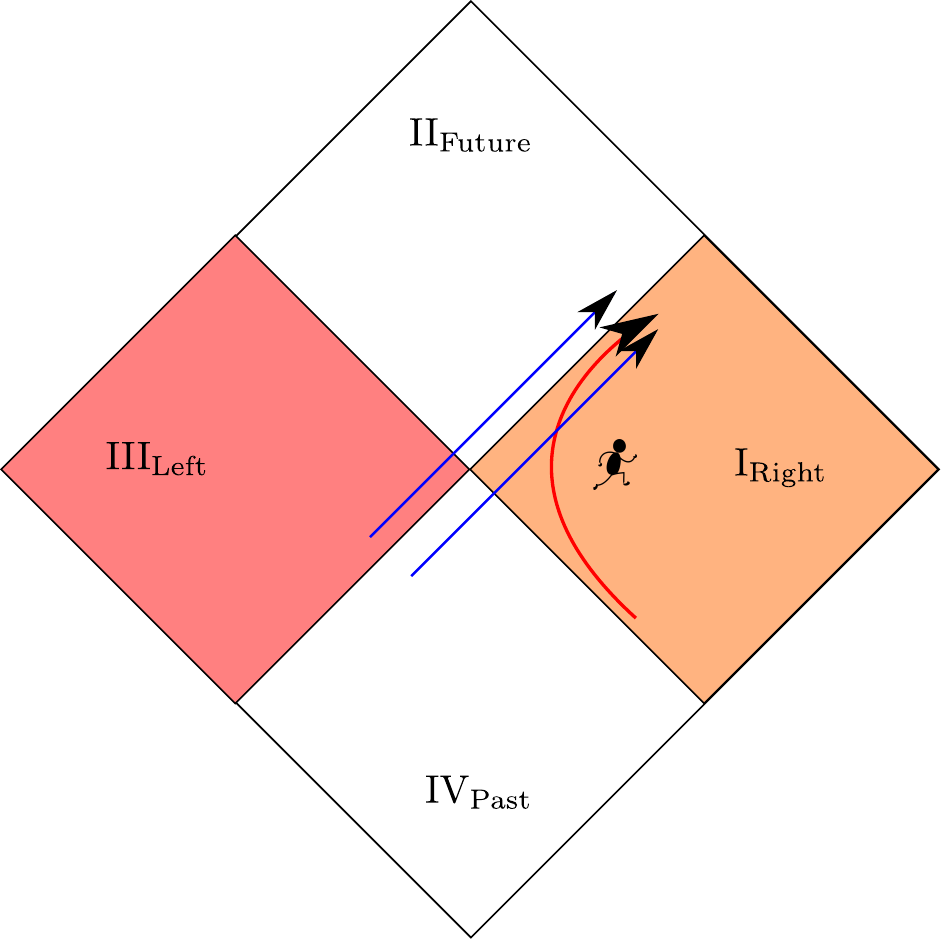} \label{NonBell}} \hspace{2cm}
\subfigure[Bell Measurement]{
\includegraphics[scale=.35]{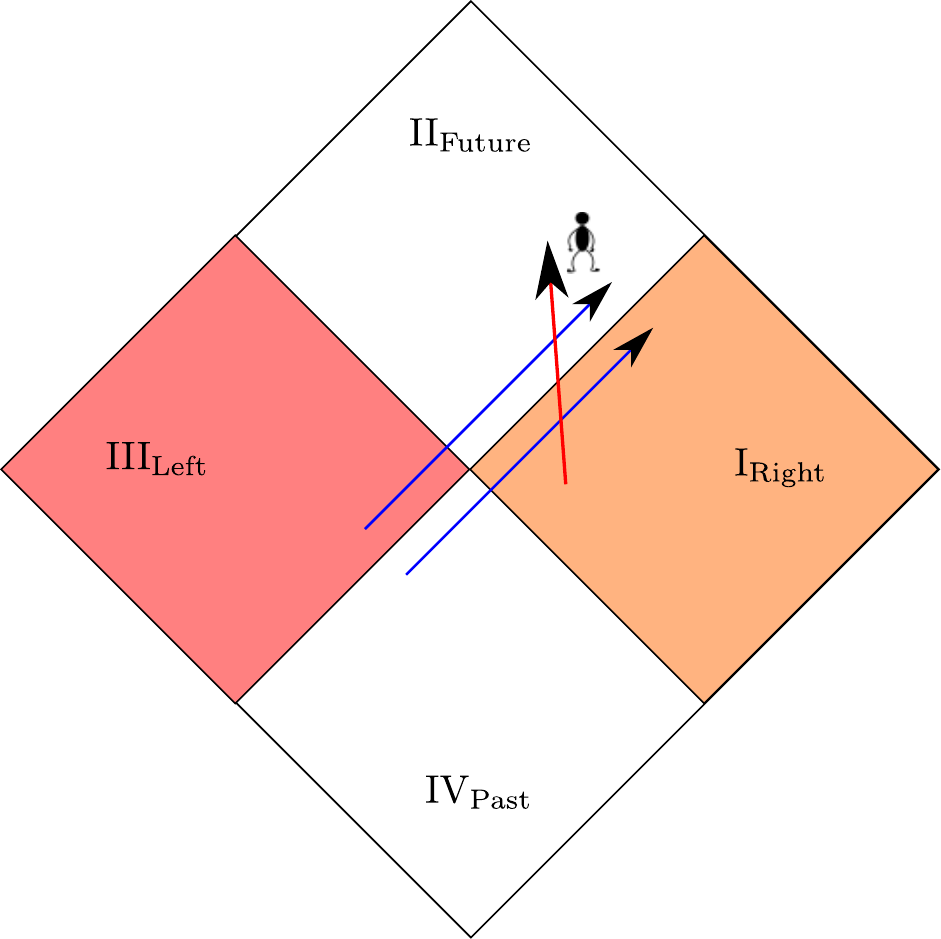} \label{Bell}} 
\caption{An accelerating observer only intersects the modes of the right wedge so can only do non-Bell measurements. These do not measure the actual state of the system but instead collapse the system into a different state. An inertial observer on the other hand intersects both modes and thus can perform a Bell measurement to verify that the full state is the Minkowski vacuum. }
\label{RindlerMinkowski}
\end{center}
\end{figure}

In light of this, our previous comment about a right moving observer finding the left movers in the vacuum only if they are in the state \bref{MinkowskiMode} can be restated in the following way. Accelerating observers who stay inside the Rindler wedge have access to only half the system can only perform non-Bell measurements and cannot tell of the full state is the Minkwoski vacuum or any other state that leaves the right wedge density matrix the same (see Figure~\ref{NonBell}). However, inertial observers can measure the full state of the system and in fact do so while crossing the horizon. They can thus tell if the full state is the Minkwoski vacuum or some other state. Thus inertial observers perform Bell measurements. This is shown in Figure~\ref{Bell}.

Now consider a CFT in a thermal state that we call system A. One can always find another system  B (which may or may not be a CFT) which together with the CFT is in a pure state. Now one can consider boundary-Alice living on system A. It is widely believed that  system A which is a  CFT captures the bulk {\em at least} outside the horizon and we will assume that. Similarly, system B can {\em approximately} capture the bulk physics external to the horizon on the other side with the approximation getting better the more system B's dynamics can be described by a CFT.

According to AdS/CFT,  boundary-Alice's interaction with the other degrees of freedom of system A  in principle describes bulk-Alice hurtling towards the horizon. However, when bulk-Alice crosses the horizon, she is performing a joint measurement on the degrees of freedom on the two sides of the horizon. The degrees of freedom inside  the horizon can be traced to modes from the other exterior region and thus are associated to degrees of freedom of system B. Thus, bulk-Alice crossing the horizon performs a joint measurement on the bulk modes associated to boundary systems A and B. Boundary-Alice on the other hand lives only on system A and cannot perform a joint measurement on the degrees of freedom of both system A and system B. This tension between the measurements accessible to bulk-Alice and boundary-Alice is may suggest that (a) system A and system B cannot capture the physics behind the horizon, or (b) the bulk physics needs modification at the horizon scale. However, since the theory of decoherence and measurements is not fully developed, especially in the context of AdS/CFT, we cannot fully rule out some surprise which can resolve the aforementioned tension between the bulk and the boundary point of view.

\bibliographystyle{jhep}
\bibliography{Final}

\end{document}